# A Computational Approach for Modeling Platelet Adhesion Dynamics on Thrombogenic Surfaces

Ali Lotfian [1, 2], Ehsan Roohi [3]

[1] Department of Mechanical Engineering, Ferdowsi University of Mashhad, P.O.Box. 91775-1111, Mashhad, Iran

[2] School of Aerospace Engineering, Xi'an Jiaotong University, 710049, Xi'an, China

[3] Mechanical and Industrial Engineering, University of Massachusetts Amherst, 160 Governors Dr., Amherst, MA 01003, USA

*Corresponding author. *E-mail addresses*: roohie@umass.edu (E. Roohi).

**Abstract**

Platelet adhesion and aggregation are essential for primary hemostasis, forming a clot that quickly stops initial bleeding. Despite this critical role, the dynamic interactions of platelet receptors with exposed collagen and von Willebrand factor (vWF) at the injury site and how these interactions influence thrombus formation under varying blood flow conditions are not fully understood. This study aimed to investigate the mechanisms of platelet adhesion and aggregation on collagen- or vWF-coated surfaces numerically. We combined the stochastic Bell's law with a deterministic elastic force featuring a time-dependent coefficient within the context of a dissipative particle dynamics (DPD) model to simulate thrombosis formation numerically. Our simulation results revealed that the numerically predicted platelet adhesion patterns closely matched experimental observations reported in the literature, demonstrating accurate replication of platelet behavior on collagen- and vWF-coated surfaces. Consequently, our deterministic/stochastic force model in DPD provides valuable insights into platelet adhesion dynamics under different flow conditions. These results contribute to a deeper understanding of platelet dynamics and potential therapeutic targets for managing hemostatic disorders.

**Keywords:** Dissipative particle dynamics, Stochastic binding, Platelet adhesion, Thrombosis, VWF, Collagen.

## 1. Introduction

Platelet adhesion is vital in preventing bleeding and pathological clot formation (thrombosis), but can also lead to pathological conditions if it becomes dysregulated. This process plays a crucial role in both hemostasis, which prevents bleeding, and thrombosis, which can result in harmful blockages within blood vessels. Excessive platelet adhesion can contribute to atherosclerosis and cardiovascular diseases, highlighting the importance of balanced platelet function. At the site of vascular injury, plasma von Willebrand factor (vWF) can become immobilized onto the exposed subendothelial surface. The interactions of adhesion receptors on platelets with vWF and exposed collagen are crucial for initial platelet adhesion. These bonds between the adhesive ligand and receptor pairs counteract the drag on the tethered platelets while platelets themselves are subjected to shear stress. This dynamic process may initiate a sequence of signaling events in platelets, activating the ligand binding function of the integrin αIIbβ3 (GPIIb/IIIa) to enhance the platelet adhesion and aggregation firmness. Thus, the dynamic interaction mediates stable platelet adhesion and aggregation, forming an irreversible stable thrombus[1,2].

Previous studies have investigated various aspects of platelet and collagen/vWF interactions [3-11]. Coarse-grained modelling, where vWF multimers and platelets are represented as interconnected particles, has revealed how flow, chain stiffness, and inter-monomer attraction govern vWF behavior under various conditions[8]. The unique "catch bond" behavior, where specific molecular bonds become stronger under mechanical stress, plays a vital role in platelet adhesion within high shear zones, and detailed studies elucidate these specific biomechanical principles at different levels of complexity[9]. Despite these insights, most coarse-grained models do not fully incorporate the complex biochemical signaling pathways that regulate platelet adhesion and activation.

Artificial intelligence (AI)-based multiscale simulations now provide unprecedented resolution, modelling millions of interacting platelets and how fluid interactions impact their function[10]. Some works analyzed the key processes driving platelets towards vessel walls (margination) and factors governing adhesive behaviors under dynamic flow conditions[11]. While AI-based models offer significant potential, their application in simulating the full spectrum of platelet dynamics, including biochemical interactions and feedback mechanisms, remains limited. Additionally, many AI-based models require extensive computational resources, which can limit their accessibility and practicality for widespread use.

Attempts have been made to develop computational biomechanics models of the platelet adhesion dynamics to elucidate the contribution of hemodynamic factors[7,12-18]. Dissipative particle dynamics (DPD) models have been used to explore platelet adhesion dynamics. Filipovic et al.[12] and Tosenberger et al.[13] used the DPD model to identify potential mechanisms underpinning why clots stop growing under flow conditions. Wang et al.[14] used this method to investigate non-physiological shear stresses associated with medical devices, potentially promoting platelet dysfunction and device-related complications. Hybrid models merging DPD with partial differential equations (PDEs) have also been proposed to understand thrombus formation. This approach can model blood flow, mechanical events, and biochemical signalling through coagulation factors alongside platelet dynamics[13]. The DPD method was employed to model the plasma flow and platelets, while the PDEs described the regulatory network of plasma coagulation. The specific PDEs used in Ref. [13] were reaction-diffusion-advection equations. These equations help in modeling the spatial and temporal changes in the concentration of coagulation factors and fibrinogen, which are critical for the formation and growth of the clot. The reaction terms account for the biochemical interactions, diffusion terms for the spread of these substances, and advection terms for their transport due to blood flow. Such models could illuminate possible factors leading to clot growth arrest under flow conditions[15].

The bonds between platelet receptors and vWF or fibrinogen do not exhibit a simple "on/off" behavior. Stochastic modelling may be needed to represent thrombosis accurately. Models that account for randomness (stochasticity) better resemble a thrombus shell's in vivo flexibility and explain a clot's unexpected disassembly behavior[16]. Despite these advancements, many models fail to fully capture platelet interactions' stochastic nature and mechanical properties under various physiological conditions. Moreover, most current models do not adequately address the role of microenvironmental factors in modulating platelet behavior and the dynamic remodeling of the thrombus.

Alongside the in vitro studies, modelling studies focused on thrombus formation have improved our understanding of how shear stress and surface interactions promote aggregation and provide critical knowledge for managing conditions involving abnormal vWF function[18]. Additionally, computational techniques have highlighted how clot contraction, where platelets actively pull on fibrin networks, can involve passively entrapped red blood cells. This can impact clotting time and clot size[7]. However, current models do not fully understand the interplay between clot contraction and the dynamic remodeling of the thrombus structure under flow conditions.

Our work introduces a multiscale platelet adhesion model that uniquely integrates dissipative particle dynamics (DPD) with Bell's stochastic binding law and a time-dependent elastic force, bridging the gaps in previous studies. Unlike Tosenberger et al. [13,19], which used DPD but did not incorporate stochastic adhesion kinetics, our model explicitly applies Bell's law to capture force-dependent platelet attachment and detachment dynamics (see Table. 1 for feature comparisons). In contrast to Kaneva et al.[17], which employed stochastic adhesion mechanics but lacked DPD and used some fluid models, our approach, which uses DPD, ensures a realistic fluid-particle interaction crucial for platelet dynamics under shear flow. Furthermore, unlike Kaneva's 2D modeling, our fully three-dimensional framework allows for more accurate thrombus formation simulations. By combining stochastic receptor-ligand interactions with a time-evolving adhesion force in a DPD-based system, our work provides an unprecedentedly detailed and biologically accurate representation of platelet adhesion and thrombus stability, making it a significant advancement over existing models.

Table. 1: Key differences between previous and present thrombus models.

| Feature | Tosenberger et al. [13,19] | Kaneva et al. [17] | Present Study |
| --- | --- | --- | --- |
| Simulation Framework | Dissipative Particle Dynamics (DPD) | Stochastic adhesion model (no DPD) | DPD + stochastic Bell's law + time-dependent elastic force |
| Receptor-Ligand Binding | Not explicitly modeled | Stochastic adhesion, but no force-dependent Bell's law for association | Explicit force-dependent Bell's law for both adhesion & detachment |
| Dimension | 3D | 2D | 3D |
| Thrombus Formation Mechanism | Based on random platelet detachment | Stochastic adhesion without hydrodynamic interaction | Dynamic platelet aggregation under shear flow with explicit force interactions |
| Force Effects on Platelets | No explicit force-dependent adhesion | Only Bell's law for dissociation | Bell's law applied to both attachment and detachment dynamics |
| Platelet Aggregation Stability | Not explicitly modeled | Limited due to 2D approach | Explicitly modeled with a time-dependent adhesion force |
| Biological Accuracy | Simplified receptor-ligand interactions | Stochastic adhesion mechanics | Stochastic receptor-ligand interaction with DPD for realistic fluid effects |

To further enhance the biological accuracy of our model, Bell's law[20] is implemented based on the rate of bond formation and rupture under external forces, allowing us to mimic receptor-ligand interactions more realistically. While Bell's model describes the behavior of a single bond, in reality, platelets adhere to the substrate through multiple receptor-ligand bonds. Our approach simplifies this process for computational efficiency while preserving the essential stochastic nature of platelet adhesion dynamics. By doing so, we strive to improve the numerical representation of thrombus formation and stability compared to in-vitro observations. This approach offers deeper insights into the biophysical mechanisms governing platelet adhesion and thrombus dynamics, ultimately contributing to the development of more effective therapeutic strategies and medical devices for thrombosis-related disorders. A detailed comparison of these features with previous models is presented in Table. 1.

## 2. Method

### 2.2 Numerical Method

#### 2.2.1 A Brief in DPD for Platelet Movement

DPD is a particle-based simulation method used for studying the dynamic and rheological properties of simple and complex fluids at the mesoscopic scale [19,21-28]. The particles follow Newton's second law, with interactions involving conservative ($\vec{F}_{ij}^C$), dissipative ($\vec{F}_{ij}^D$), and random ($\vec{F}_{ij}^R$) forces [23,29]. The governing equations for DPD particle motion are:

$$\frac{d\vec{r}_i}{dt} = \vec{v}_i, \qquad \frac{d\vec{v}_i}{dt} = \frac{1}{m_i}\sum_{j\neq i}(\vec{F}_{ij}^C + \vec{F}_{ij}^D + \vec{F}_{ij}^R) \qquad (1)$$

Here, $m_i$ represents the mass, $\vec{r}_i$ denotes the position vector, and $\vec{v}_i$ represents the velocity vector of particle $i$. The summation is taken over all interacting particles $j$ accounting for the influence of every other particle in the system. The force components are defined as [14]:

$$\vec{F}_{ij}^C = \omega_c(r_{ij})\ \vec{e}_{ij} \qquad (2)$$

$$\vec{F}_{ij}^D = -\gamma\ \omega_D(r_{ij})\ (\vec{e}_{ij}\cdot\vec{v}_{ij})\vec{e}_{ij} \qquad (3)$$

$$\vec{F}_{ij}^R = \varphi\ \omega_R(r_{ij})\ \theta_{ij}\vec{e}_{ij} \qquad (4)$$

, where $\omega_C(r_{ij})$ is the conservative weight function, $\gamma$ and $\varphi$ are coefficients for the dissipative and random force ($\varphi = \sqrt{2\gamma k_B T}$ where $k_B$ is the Boltzmann constant and T is the temperature [25]), $\omega_D(r_{ij})$ is a weighting function, $\varphi$ is a noise amplitude parameter, $\omega_R(r_{ij})$ is the weighting function for the random force $\theta_{ij}$ is a normally distributed random variable with zero mean ($\theta_{ij} = \theta_{ji}$). The distance between particle $i$ and $j$ is denoted as $r_{ij} = |\vec{r}_{ij}| = |\vec{r}_i - \vec{r}_j|$, where $\vec{r}_{ij}$ represents the position vector from particle $j$ to particle $i$. Additionally, $\vec{v}_{ij}$ represents the relative velocity between particles $i$ and $j$, while $\vec{e}_{ij} = \vec{r}_{ij}/|r_{ij}|$ the unit vector pointing from particle $j$ to particle $i$. DPD simulations use a modified velocity-Verlet (vV) algorithm to integrate the equations of motion [23,30,31], which consists of the following steps. First, the half-step velocities are computed using:

$$v_i^{n+1/2} = v_i^n + \frac{1}{2}a_i^n\Delta t, \qquad (5)$$

where $n$ is the time step index, representing the current step in the simulation; $a_i^n$ is the acceleration of particle $i$ at time step $n$, calculated from the forces acting on the particle. Next, the positions are updated according to:

$$r_i^{n+1} = r_i^n + v_i^{n+1/2}\Delta t \qquad (6)$$

Once the positions are updated, new accelerations are computed using the updated forces. Finally, the full-step velocities are obtained as:

$$v_i^{n+1} = v_i^{n+1/2} + \frac{1}{2} a_i^{n+1} \Delta t \qquad (7)$$

This method ensures stability and time reversibility, making it suitable for mesoscopic simulations.

Blood was modeled as a composite consisting of platelet particles and plasma particles. Inactive platelets exhibit a behavior akin to plasma particles, except for their ability to undergo activation and adhere to protein-coated surfaces. Specifically, the interactions involving platelets with plasma particles or between two inactive platelets adhere to the principles delineated by Eqs. 1–4. Meanwhile, the interactions between platelets and protein particles conform to an elastic model adapted from the studies by Tosenberger et al.[19], Wang et al.[14] and Liu et al.[32], as well as a stochastic spring force model of Ref. [33].

**2.2.2 Validation of DPD solver for pressure-driven (Poiseuille) flow**

To validate the DPD solver in LAMMPS, the velocity profile in Poiseuille flow obtained from the DPD solution in this study was compared with the MD and DPD results from Keaveny et al.[34]. The studied test case is the flow in a straight channel of the size $34.2\sigma \times 34.2\sigma \times 8.85\sigma$. The fluid density is $0.8\ \sigma^{-3}$. Periodic boundary conditions are applied in $x$ and $z$ directions. In the $y$ direction, the fluid is bounded by side walls that have the same density as the fluid. Each side wall consists of two layers of equally spaced particles. To prevent the penetration of fluid particles into the walls, a bounce back condition is applied at the solid-fluid interface. Different values of coarse-graining parameter were studied in Ref [34] among which we choose the value of 3 which had an acceptable agreement with the MD results. To induce the flow in the nanochannel, the force $0.0273\ m\sigma/\tau^2$ is applied to each fluid particle.

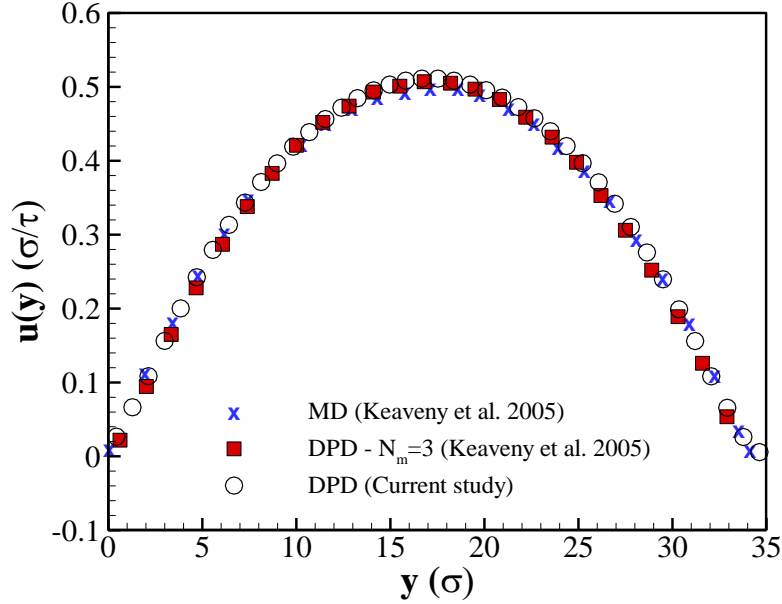

**Fig. 1:** Velocity profiles of the Poiseuille flow simulation. The velocity data is presented in dimensionless units for comparison across different methods. The blue crosses represent MD data from Ref. [34], the red squares correspond to DPD simulations with $N_m = 3$ from Ref. [34], and the open circles indicate results obtained from LAMMPS in the present study.

The width of the channel (in the $y$ direction) is divided into equal bins, and the average velocity in each bin is calculated by sampling the particle velocities. The variation of $x$-component of velocity vector across the channel width is presented in Fig. 1 and compared with the results of [34]. The good agreement of results indicated the accuracy of DPD implementation in the LAMMPS package and the suitability of the simulation setup.

**2.2.3 Particles Interaction**

The interaction between platelets and plasma particles is described using the same DPD force models described in Section 2.2.1. In contrast, platelet-collagen interactions are modeled using two main deterministic force components: the elastic force $\vec{F}_{ij}^E$ and the viscous forces $\vec{F}_{ij}^V$ between particles $i$ and $j$ are formulated as follows[19]:

$$\vec{F}_{ij}^E = a\left(1 - \frac{r_{ij}}{r_c}\right)\vec{e}_{ij}, \quad \text{where} \quad a = \begin{cases} -a_0, & \text{when } r_c \geq r_{ij} > r_a \\ a_1, & \text{when } r_{ij} \leq r_a \end{cases} \quad (8)$$

$$\vec{F}_{ij}^V = -\gamma\left(1 - \frac{r_{ij}}{r_c}\right)^2 (\vec{e}_{ij}\cdot\vec{v}_{ij})\vec{e}_{ij}, \quad \text{when } r_{ij} \leq r_c \quad (9)$$

In these equations, $a_0$ and $a_1$ represent activation state-dependent force coefficients, which will be explained later. $r_a$ is the distance at which the attraction force between two particles becomes repulsive to prevent particle overlapping. In the current work, $r_a = 2r_c/3$, is the effective distance of interaction.

Wang et al.[14] proposed a three-phase model for platelet activation, as illustrated in Fig. 2. In this model, platelets are initially inactive (green). When an inactive platelet reaches a critical distance $r_c$ from an active platelet, it transitions to Phase 1 activation (blue) within $t_1$, which is set to 0.01 of the DPD characteristic time (approximately 0.16 ms). If it remains in proximity to an active platelet for $t_2$, which is set to 300 times the DPD characteristic time (approximately 4.76 s), it progresses to Phase 2 activation (red). According to Wang et al.[14], platelet interactions evolve through the aggregation initiation phase, during which force coefficients change with activation, ultimately leading to the final aggregation phase, where fully activated platelets form stable bonds. Table 2 reports the coefficients related to the platelet–platelet interaction force.

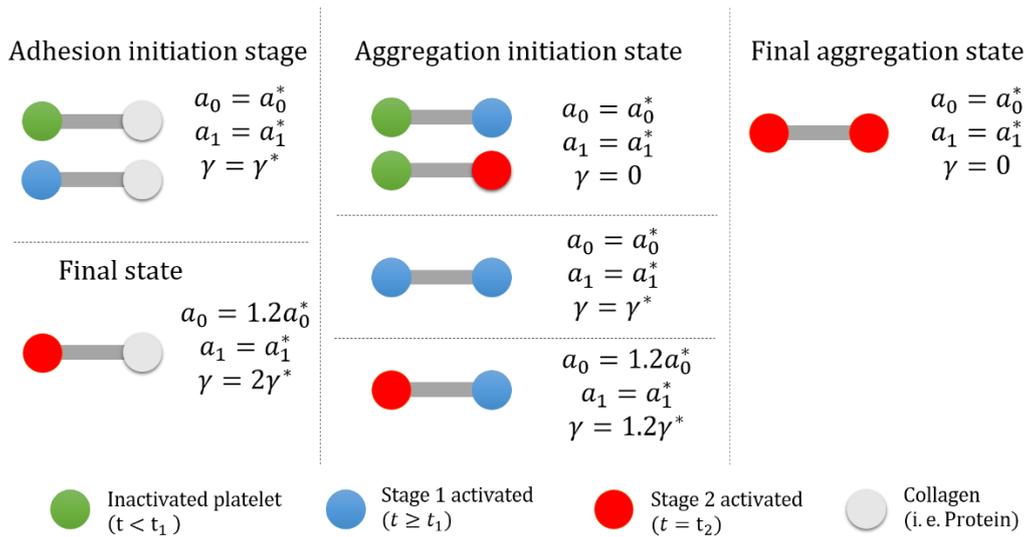

**Fig. 2:** Factors adjusting the force coefficients between platelets and platelets at different activation stages. Redrawn and adapted from Wang et al. [14] with permission (Springer, License no. 5995940965683).

**Table 2:** Force coefficients for Platelet-Platelet.

| $a_0$ | $\gamma$ | $a_1$ | $r_c$ |
|---|---|---|---|
| 10 | 4.5 | 120 | 1 |

Platelet-collagen interactions are modeled using two main deterministic force components: the elastic force $\vec{F}_{ij}^E$ and the viscous forces $\vec{F}_{ij}^V$ between particles $i$ and $j$ are formulated as follows[19]:

$$\vec{F}_{ij}^E = \begin{cases} a\left(1 - \dfrac{r_{ij}}{r_a}\right)\vec{e}_{ij} & ; r_{ij} \leq r_c \\ 0 & ; r_{ij} > r_c \end{cases}, \quad \text{where} \quad a = \begin{cases} a_t t + a_0 & ; q_{ij} > 0 \\ 0 & ; q_{ij} = 0 \end{cases} \quad (10)$$

$$\vec{F}_{ij}^V = -\gamma\left(1 - \dfrac{r_{ij}}{r_c}\right)^2 (\vec{e}_{ij} \cdot \vec{v}_{ij})\vec{e}_{ij}, \quad \text{when } r_{ij} \leq r_c \quad (11)$$

, where $q_{ij}$ is a parameter indicating the activation stage of the platelet $a_0$ and $a_t$ and are non-temporal and temporal term force coefficients. When the distance between the platelet and collagen-vWF is reduced to $r_c$, the platelet may adhere to the protein particles or aggregate with other activated platelets. $r_a/r_c$ represents the distance at which the attractive force switches to repulsion relative to the cutoff radius. This parameter is crucial in preventing platelet overlap by defining the range where attraction transitions to repulsion. $r_c$ values are given in the caption of the figures of results. The schematic of the platelet-collagen force is shown in Fig. 3.

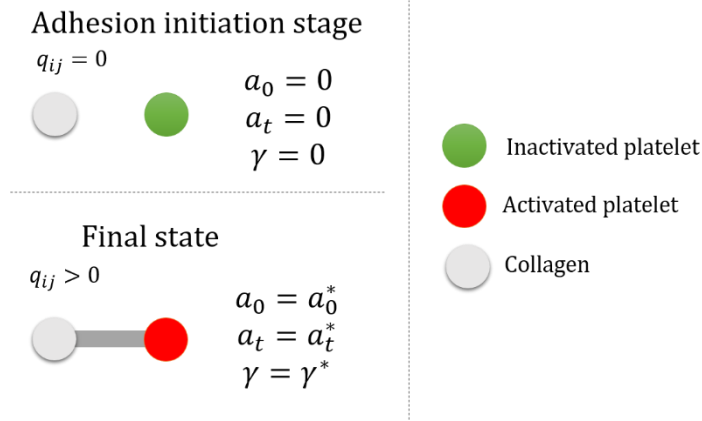

**Fig. 3:** Force between collagen and platelets (present model).

Since this model doesn't explicitly include receptor binding kinetics, we consider the evolution of adhesion force over time. In the simulations, inactivated platelets stayed within the cut-off radius of collagen particles longer than $t_1$ entered the initial activation stage ($q_{ij} > 0$). Platelets adhere to collagen either through GPIb-vWF-collagen or through α2β1 integrin-collagen bonds. As we do not consider the number of bonds per platelet, this does not have an impact on our computational model[14].

**2.2.4 Bell Law (Probabilistic)**

To the best of our knowledge, reported DPD simulations did not match the recent in-vitro observations of the clot formations. The recent in-vitro showed that scattered clots form on the surface covered by protein. Therefore, not all the surface is covered by clots. To be able to capture this scattered clot formation, a probabilistic function should be considered in addition to the deterministic force. To further support this idea, we refer to the work of Kaneva et al.[17], which used a complex stochastic spring model to improve the accuracy of the platelet adhesion to the collagen in numerical simulations. Here, to assess the initial adhesion and de-attachment of inactivated platelets on collagen and vWF, the stochastic Bell's law model[20] is implemented. The stochastic Bell's law model is a mathematical framework used to describe the kinetics of bond formation and dissociation under the influence of mechanical forces. It extends the classical Bell's model by incorporating stochastic (random) elements to account for the variability observed in biological systems.

The stochastic spring model describes the possibility of deactivation of the initially attached platelets. At the same time, it evaluates the chance of activation of inactivated platelets. The model simulates platelet interactions with proteins. The probability of an event occurring during the current time step is denoted by $P$. A uniformly distributed random number $R_{nd}$ ($0 \leq R_{nd} \leq 1$) is generated. If $R_{nd} < P$, then an event occurs; otherwise, it does not. This stochastic model allows us to depict single-platelet interactions with collagen and vWF, the plasticity of the thrombus, and its shape.

This model represents platelet bonds with collagen/vWF sites on the conduit wall as linear springs. In this model, the spring coefficient is given by Ref. [33]:

$$k(F) = \begin{cases} k_0 & \text{attach} \\ k_0 \exp(F/F_0) & \text{detach} \end{cases} \quad (10)$$

The equation describes $k_0$ as the rate of attachment and detachment solely due to thermal fluctuations at zero force, while $F_0$ is defined as $k_B T/\xi$, representing the characteristic bond force concerning the Boltzmann constant ($k_B$), temperature ($T$), and characteristic bond separation distance ($\xi$).

The Bell model can analyze crosslink detachment under a constant force $F$. This model computes the probability $p(F, t)$, indicating the likelihood that a crosslink exposed to force $F$ will detach at a given time $t$. The Bell model incorporates the effects of force on bond dissociation, providing a framework to predict how forces influence the stability and lifetime of molecular interactions, such as those seen in platelet adhesion and thrombus formation.

$$p(F, t) = k(F) \exp(-k(F) t) \quad (11)$$

The anticipated attachment duration of a solitary crosslink, denoted as $\tau$ can be calculated as follows:

$$\tau = \int_0^\infty \bar{t} \, p(F, \bar{t}) d\bar{t} = \frac{1}{k(F)} \quad (12)$$

In our simulations, we must determine the probability, denoted as $P$, of a crosslink detaching within a specific time interval between $t = t_0$ and $t = t_0 + \Delta t$. Here, $t_0$ represents the current time in the simulation, and $\Delta t$ represents the numerical time step. This probability is derived from Eq. (15) as follows:

$$p(F, \Delta t) = \frac{\int_{t_0}^{t_0+\Delta t} p(F, \bar{t}) d\bar{t}}{\int_{t_0}^{\infty} p(F, \bar{t}) d\bar{t}} = 1 - exp\left(-k(F)\Delta t\right) \qquad (13)$$

The selection of $p(F, t)$ is driven by Eq. (15), where $p(F, \Delta t)$ displays independence from $t_0$. This signifies that the detachment probability remains unaffected by historical factors, a characteristic indicative of random thermal fluctuations. Substituting Eq. (12) in Eq. (15) yields:

$$P_{as} = 1 - exp\left(-k_0 \Delta t\right) \qquad (16)$$

$$P_{dis} = 1 - exp\left(-k_0 \exp\left(F/F_0\right) \Delta t\right) \qquad (17)$$

The force employed in Eq. (17) is the external force given by Eqs. (2-4).

Fig. 4 shows the schematic diagram of the platelet adhesion process and clot formation under the influence of blood flow. The platelets are depicted as red spheres, indicating their activation in forming bonds with the surface (collagen or von Willebrand factor). The bonds formed between the platelets and the surface are represented by blue and red circles, where blue indicates bond creation ($P_{as}$) and red represents bond detachment ($P_{dis}$). Additionally, the effect of the external force on bond dissociation is shown, which plays a crucial role in breaking the bonds and altering them due to the impact of blood flow. The surface of collagen or VWF is simplified as gray spheres, although, in reality, these proteins form a complex structure on the surface. On the left side of the diagram, the aggregation of platelets forms the "Big Clot," which results from repeated platelet binding with the surface. The external force is applied in Eq. (17). According to the probability of association and dissociation, the binding status will be modified:

$$\text{if } P_{as} > R_{nd} \quad \text{Binding created} \qquad (18)$$

$$\text{if } P_{dis} > R_{nd} \quad \text{Binding disscoaited} \qquad (19)$$

Bell's law, combined with the time-dependent elastic force model of Eq. (10), sets up a probabilistic-deterministic force model to mimic in-vivo adhesion of the platelets to the protein-covered surface.

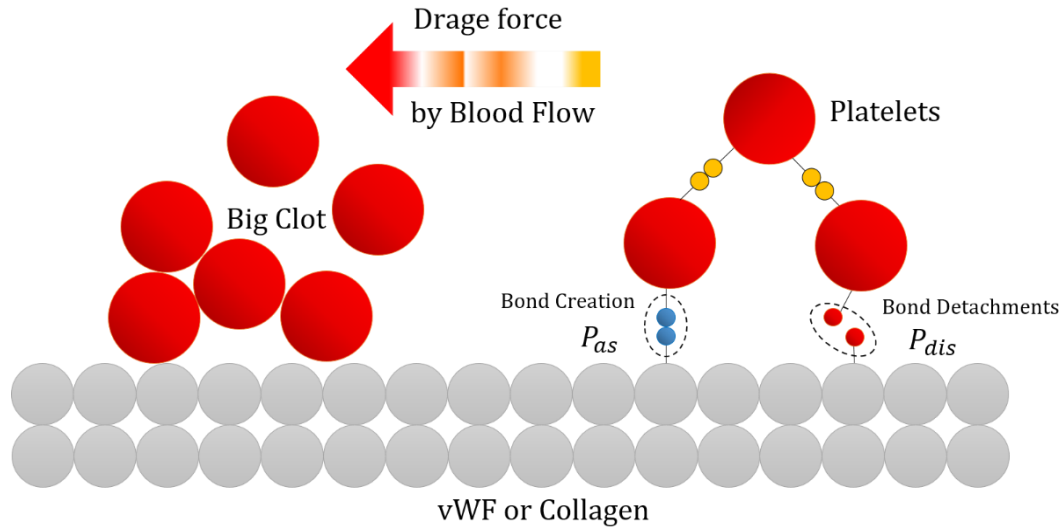

**Fig. 4:** Schematic diagram illustrating platelet adhesion and clot formation under the influence of blood flow, including bond creation, detachment, and the effect of external force on bond dissociation.

**2.3 Simulation Algorithm in LAMMPS**

Fig. 5 shows the simulation algorithm procedure implemented in the LAMMPS. The rectangle indicated by dashed lines on the left shows the changes considering the Bells' law implementations. In addition to Bell's law, we modified the DPD force models in LAMMPS to create a time/activation-dependent elastic force between platelets and collagen/vWF.

Several significant changes have been made to LAMMPS's functions. These include the addition of the Bell's model based on the external DPD force and the creation of a time/activation dependent force between platelets and protein-covered surface. These modifications have been instrumental in our ability to accurately simulate platelet behavior and its interaction with the surrounding environment.

1- We added a function called pair_plafn. This code implements the stages of activation of the platelets.
2- Changes in pair_dpd code: This function is already available in LAMMPS but it was extensively modified in the current work:
   2-1- We created a list of platelets. The original code had a list of all particles used in the simulations for all types in one list.
   2-2- We have defined an array to store the DPD force from the plasmas to the platelets.
   2-3- During the DPD force calculation algorithm (which is already available in the original code), if we encounter a pair of plasma-platelet particles, we add the force between them to the platelet DPD force array.

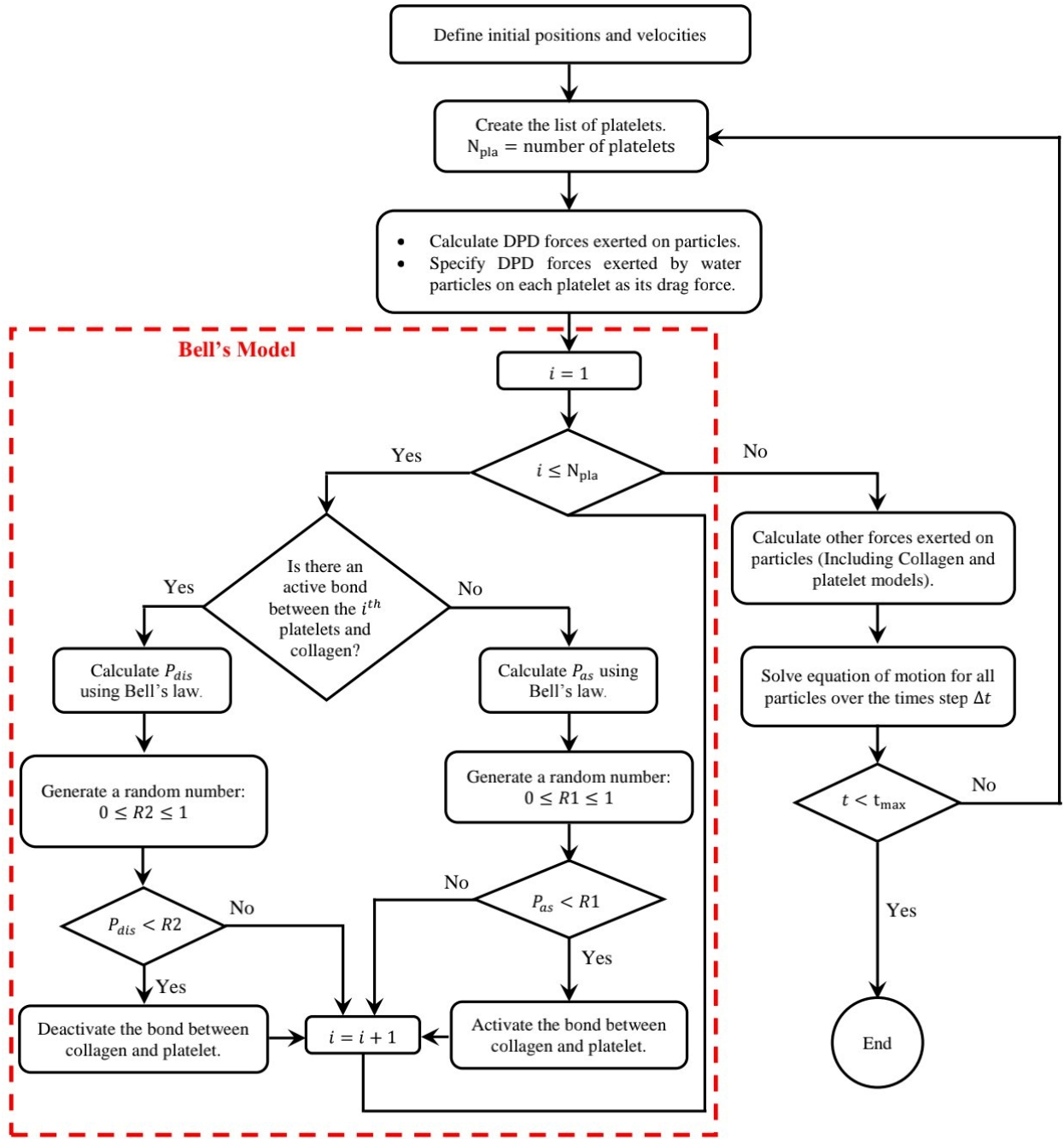

**Fig. 5:** DPD-Bell's law flowchart.

2-4- We calculated the probability of activation for each inactive platelet with the help of Bell's law, and with an acceptance-rejection method, we activated the desired platelet with the calculated $P_{as}$ probability (Eq. 16).

2-5- We calculated the probability of deactivation using Bell's model (based on the external drag force) for each active platelet. With an acceptance-rejection method, we deactivated the desired platelet with the calculated $P_{dis}$ probability (Eq. 17).

2-6- We created a time/activation elastic force between platelets/proteins based on Eqs. (16-17).

We conducted simulations of the adhesion of platelets on collagen/vWF within a three-dimensional channel. The dimensions of the channel were set at $40 \times 10 \times 10\ l^3$ DPD (equivalent to $200 \times 50 \times 50\ \mu m^3$) as illustrated in Fig. 6. The platelet concentration was maintained at a constant value of $2\times10^6$ platelets/mm$^3$. This relatively elevated platelet concentration was intentionally chosen to expedite the clotting process. Increasing the number of platelets enhances the interactions between platelets and the injured site, leading to faster clot formation. This approach allows for a more efficient investigation of the mechanisms involved in platelet adhesion, aggregation, and thrombus growth under various experimental conditions. The bottom wall comprised layers of collagen or vWF particles (yellow region in Fig. 6), replicating the injury site on the blood vessel. A consistent horizontal body force directed towards the outlet mimics the pressure that propels the flow from the inlet to the outlet. The simulation covered approximately 60 seconds of simulated time. This setup replicates the conditions necessary to study platelet adhesion, aggregation, and thrombus formation under a controlled flow, allowing for an accurate analysis of the dynamic processes involved in clot formation and the effects of varying flow conditions on these processes.

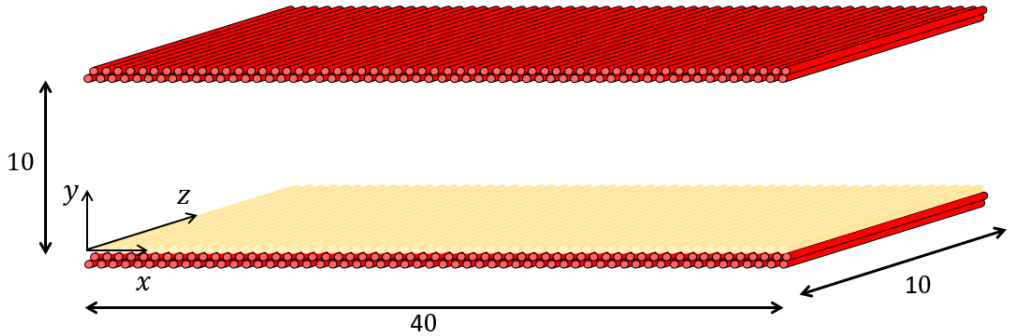

**Fig. 6:** DPD Flow domain of the micro-channel with a bottom wall covered with collagen or vWF (yellow region). The dimensions are in DPD units.

## 3. Results

### 3.1 Single Platelet Trajectory

To accurately assess the implementation of Bell's law, we analyzed the behavior of a single platelet situated at a specific location on the bottom wall under three distinct scenarios: full association ($P_{as} = 1, P_{dis} = 0$), full dissociation ($P_{as} = 0, P_{dis} = 1$), and conditions dictated by Bell's law, where the probabilities of association ($P_{as}$) and dissociation ($P_{dis}$) depend on the specific parameters $k_0$ and $F_0$, as illustrated in the Fig. 7, which shows the effect of different adhesion and repulsion coefficients on the motion of a single platelet along the dimensionless channel length, i.e., $X/L$. The study was conducted at two different flow speeds: 0.42 mm/s and 0.82 mm/s. In the case of full association, the platelet remains relatively stationary with only minor oscillations, irrespective of the flow speed. Under Bell's law condition (frame a), the platelet exhibits a stop-and-go movement until it nears the end of the channel. Conversely, the platelet under full dissociation progresses more swiftly towards the channel's exit, leaving the channel around $t = 7$ s. At the higher flow speed (frame b), the behaviors of the platelets under both full dissociation and Bell's law conditions appear more similar, with both reaching the channel's end in less than 5 seconds. This observation underscores the significant impact of the flow's shear rate on the adhesive dynamics of the platelets, demonstrating that it is a critical factor alongside Bell's law in determining platelet adhesion behavior.

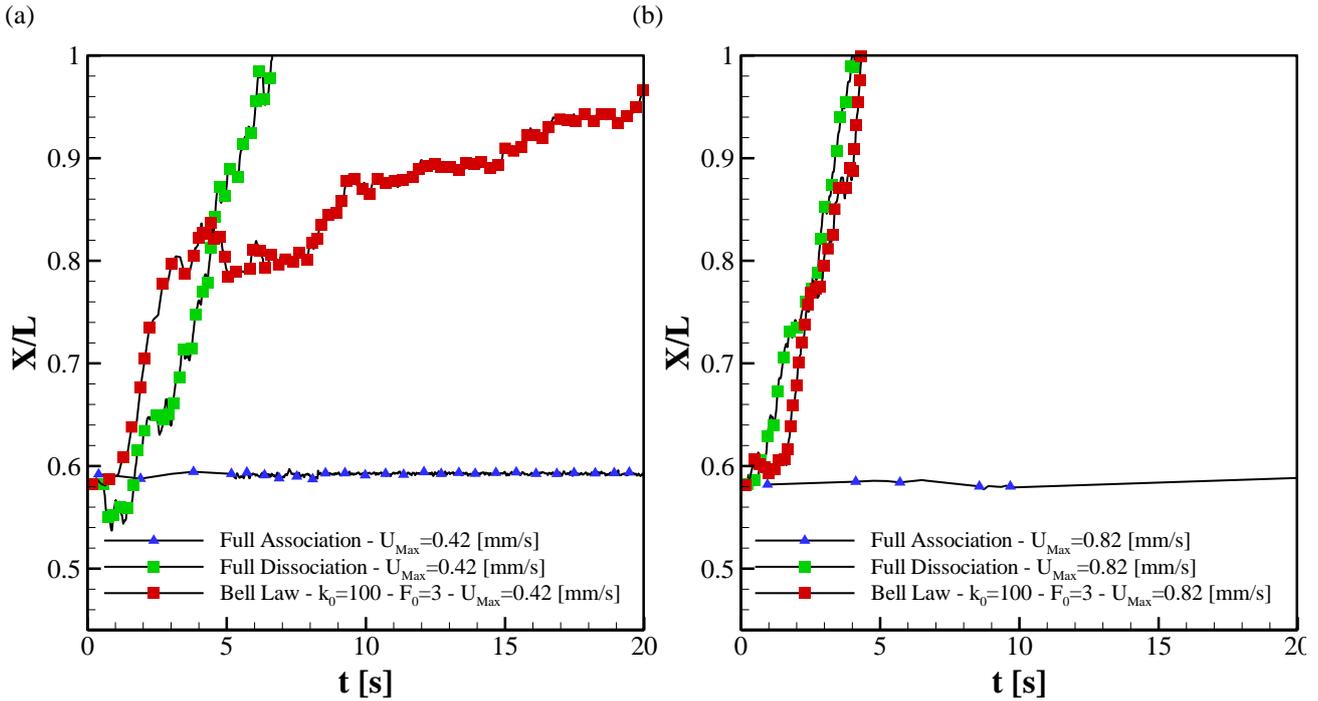

**Fig. 7:** Temporal variation of the trajectory of a single platelet at different conditions.

**3.2 Validation of the code for clot formed over a small region covered by collagen**

To delve deeper into the practical application of Bell's law, our study incorporated a channel with its bottom surface partially covered in collagen. In Fig. 8, three distinct frames highlight different aspects of platelet adhesion dynamics under varying conditions. Frame (a) demonstrates a scenario where a thin blood clot has formed over the collagen-covered surface, calculated using specific $k_0$ and $F_0$ coefficients, resulting in probabilities of association ($P_{as}$) of 0.503 and dissociation ($P_{dis}$) of 0.907. This frame illustrates a situation where both association and dissociation processes are significantly active, contributing to the dynamic equilibrium of the clot formation. Frame (b) focuses on the full association condition, where $P_{as} = 1$ and $P_{dis} = 0$. This condition supports platelet attachment with occasional detachment, leading to clot re-growth. The outcomes observed here align with those documented in References [5,19], indicating an adhesion/detachment scenario. However, this repetitive growth/collapse was not observed in the experiments reported in the literature. We should remember that with consideration of a time-dependent elastic force coefficient, we were able to remove this behavior from DPD simulations. Even though our model uses a linear increase in force coefficient, the employed detachment rate in Bell's law uses the same force and this helps in avoiding formation of unphysical results, i.e., big clots. If we do not use a time-dependent force coefficient, the simulation will result like Fig. 8, top-right frame, where clots will be formed and detached, which is not physical. Our numerical experiment showed that we need this time-dependent force coefficient to capture simulation results closer to experimental observations that show scattered clots.

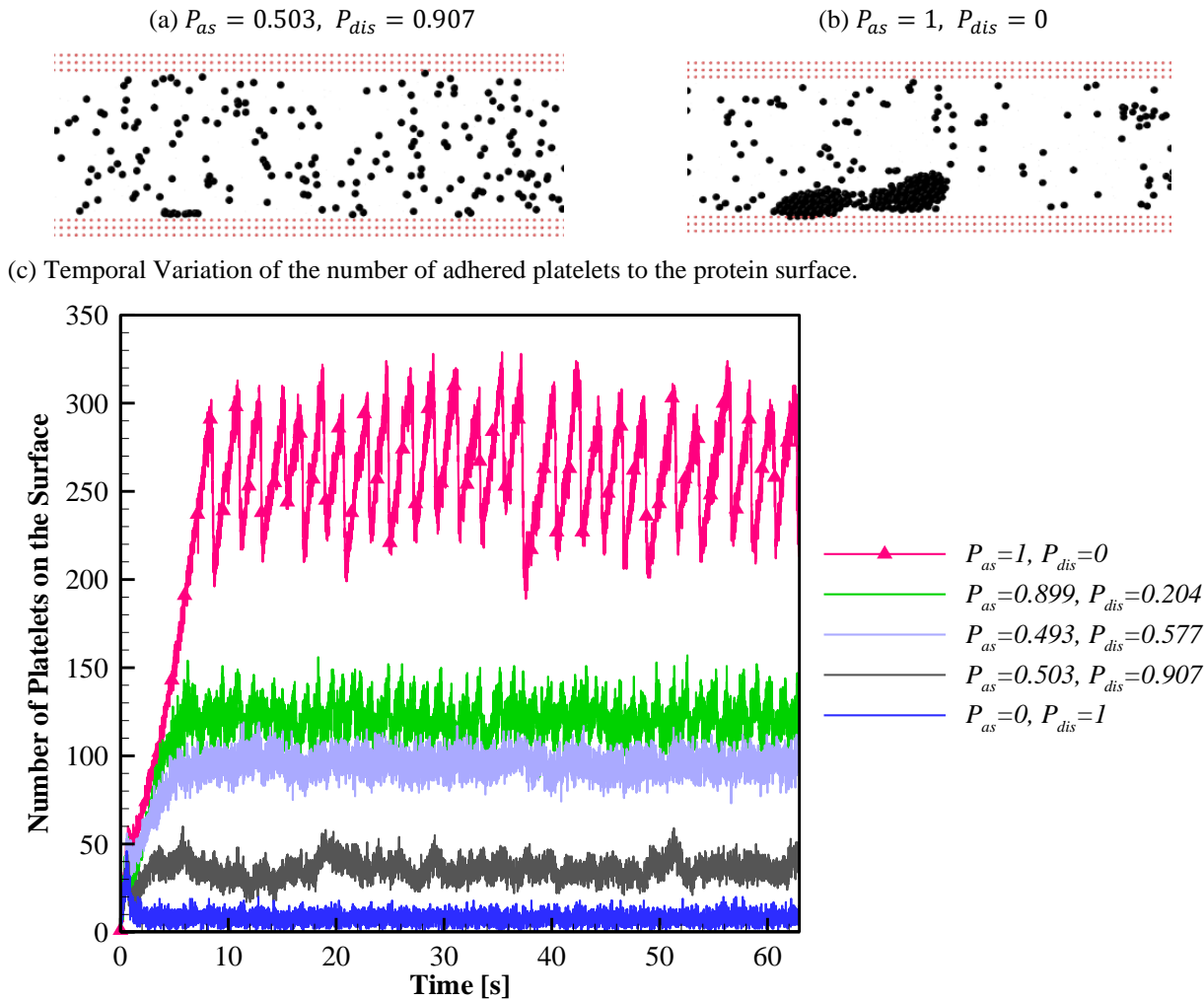

(a) $P_{as} = 0.503$, $P_{dis} = 0.907$　　　　(b) $P_{as} = 1$, $P_{dis} = 0$

(c) Temporal Variation of the number of adhered platelets to the protein surface.

**Fig. 8:** Clot formation and detachment at various dissociation association rates.

Frame (c) details the temporal variations in the number of platelets adhering to the wall under different association and dissociation conditions. In the full association case, the graph shows periodic fluctuations in the number of adhered platelets, depicting a decline followed by an increase indicative of the ongoing clot detachment and regrowth processes. The full dissociation case is marked by a near absence of platelet adhesion over time, highlighting the extreme of the dissociation dominance. Other cases, influenced by varying $P_{as}$ and $P_{dis}$ values, demonstrate intermediate behaviors between these extremes. Overall, Fig. 8 compellingly illustrates that Bell's law enables the modelling of diverse trends in platelet adhesion to the surface, capturing a spectrum of physiological responses under different biochemical and biomechanical conditions. This aids in understanding the complex dynamics of clot formation and stability under various physiological conditions.

### 3.3 Numerical Parameter Study of Platelet-collagen adhesion

Fig. 9 shows different clot formation patterns obtained using various numerical parameters. Different clots can be obtained by changing the numerical parameters in Bell's law. The clot formation patterns obtained from DPD simulations exhibit a clear dependence on adhesion parameters, with increasing adhesion strength leading to more pronounced and structured clot formations. At low adhesion values ($a = 0$), clotting remains sparse, with only small and scattered aggregates forming on the surface. As the adhesion parameter increases ($a = 10, 100$), larger clusters emerge, indicating enhanced particle binding and aggregation. At the highest adhesion values ($a = 900$), extensive clot formation is observed, with dense, well-defined structures covering a significant portion of the simulated domain. The impact of inlet modifications is also evident, further influencing clot distribution by potentially altering flow conditions and particle interactions. Comparison with experimental fluorescent microscopy images from[35], which visualize platelet aggregation and plasma adhesion to the vein surface, reveals similar clustering behavior. The experimental images display distinct, non-uniform adhesion patterns, where clotting occurs in discrete patches rather than forming a continuous layer. This aligns well with the numerical results, where adhesion strength dictates the extent of aggregation. The observed agreement between numerical and experimental findings suggests that Bell's law-based adhesion modeling effectively captures key clotting dynamics.

(a) DPD: $k_0 = 200, F_0 = 2, a_t = 0, a_0 = 1, r_c = 0.67$

(b) DPD: $k_0 = 200, F_0 = 2, a_t = 10, a_0 = 1, r_c = 0.67$

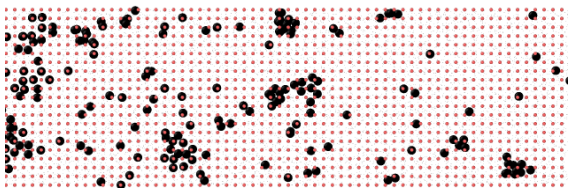
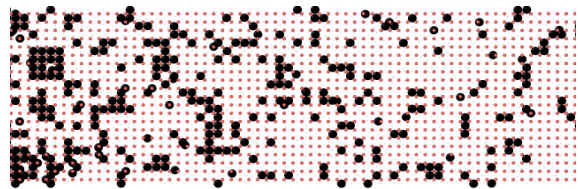

(c) DPD: $k_0 = 200, F_0 = 2, a_t = 100, a_0 = 1, r_c = 0.67$

(d) DPD: $k_0 = 200, F_0 = 2, a_t = 900, a_0 = 1, r_c = 0.67$ (inlet modified)

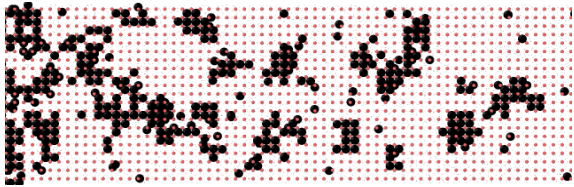
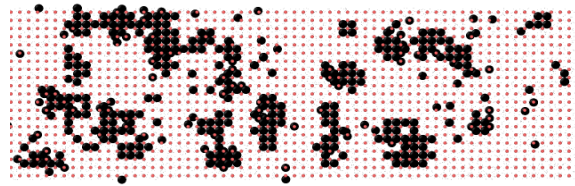

(e) DPD: $k_0 = 200, F_0 = 2,$

(f) Experimental data from Ref. [35] with permission.

$a_t = 900, a_0 = 1, r_c = 0.67$

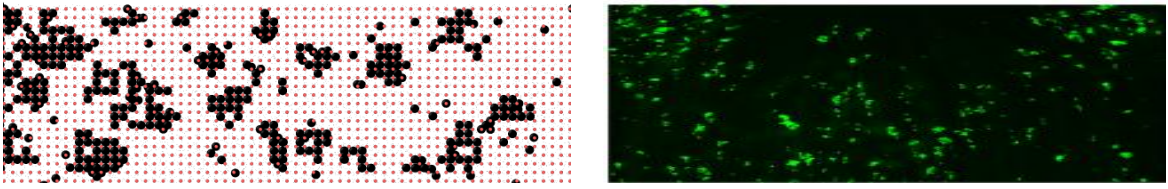

**Fig. 9:** Clot patterns at different numerical parameters, a-e) current numerical, f) experimental data from Ref. [35], reprinted with permission from John Wiley and Sons (License no: 5995941473605).

Fig. 10 illustrates the pattern of platelets adhered to the surface using detecting polygons. To achieve this, we used a C++ code to read the positions of adhered platelets and arranged them in close proximity, forming distinctive polygons. The patterns recognized by our algorithm show the distinctive formation of clots over the collagen surface under various simulation conditions. Frames corresponding to smaller $a_t$ coefficients show scattered clots on various random locations on the surface. Frame (e), which corresponds to the full association case, shows various clots of different sizes over the surface.

(a) DPD: $k_0 = 200$, $F_0 = 2$, $a = 1$, $a_t = 90$, $a_0 = 1$, $r_c = 0.67$

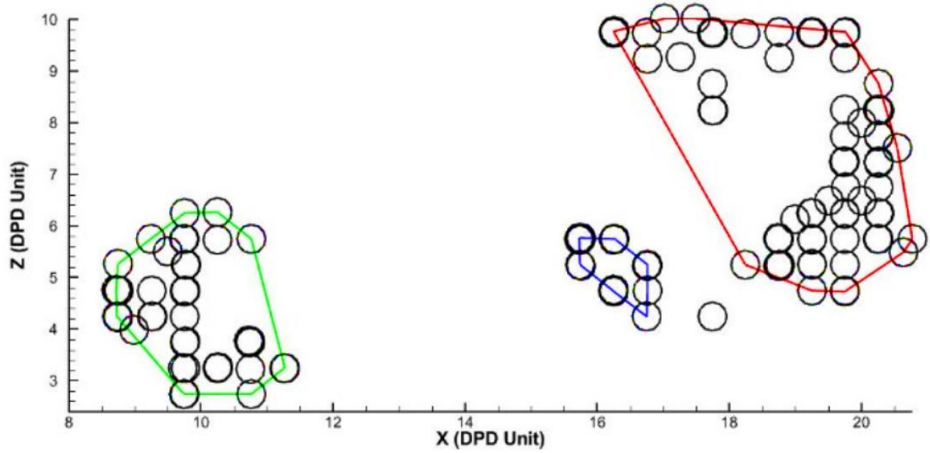

(b) DPD: $k_0 = 200$, $F_0 = 2$, $a = 1$, $a_t = 500$, $a_0 = 1$, $r_c = 0.67$

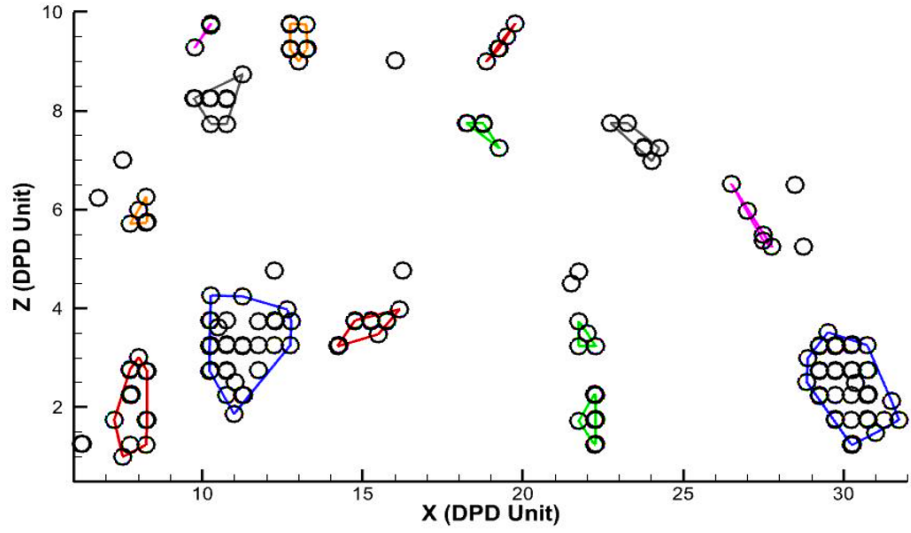

(c) DPD: $k_0 = 200$, $F_0 = 2$, $a = 1$, $a_t = 600$, $a_0 = 1$, $r_c = 0.67$

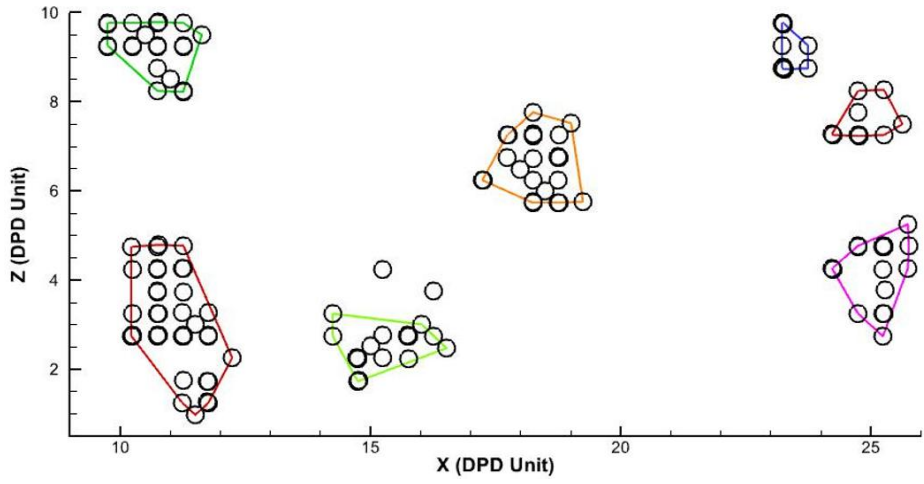

(d) DPD: $k_0 = 200$, $F_0 = 2$, $a = 1$, $a_t = 700$, $a_0 = 1$, $r_c = 0.67$ (inlet modified)

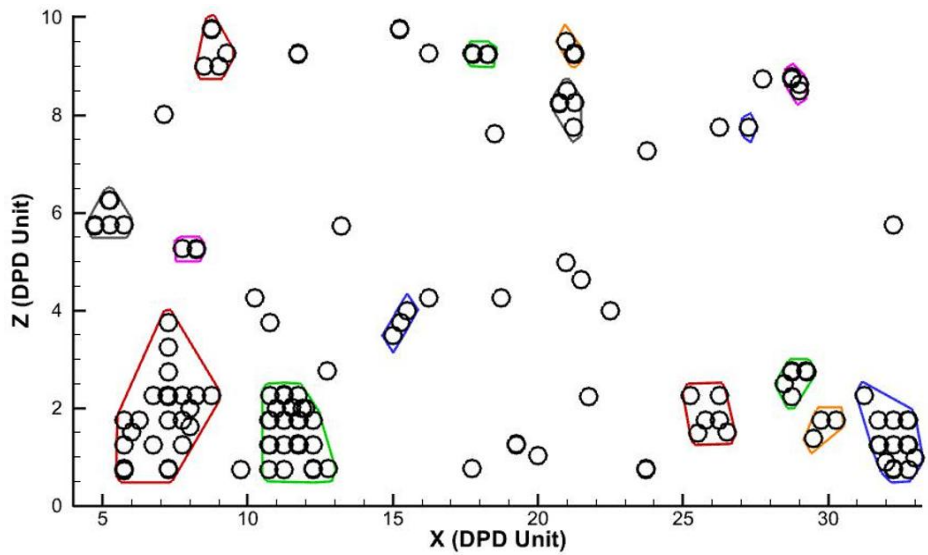

(e) Full Association Case

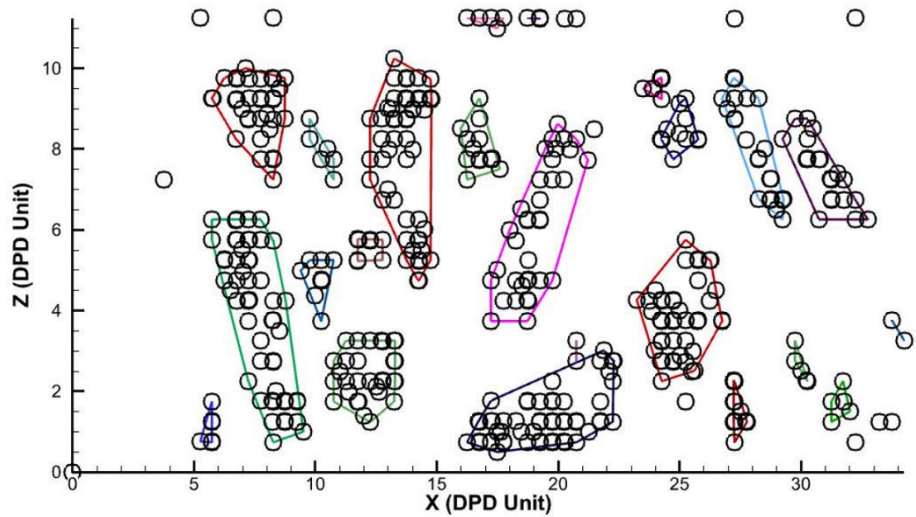

**Fig. 10:** Pattern recognition of the clot size for different cases.

Fig. 11 presents a bar chart illustrating the percentage of activated platelets with time. Activation occurs when the association probability exceeds a random threshold; at the same time, the dissociation rate is below another random number. The data shows a progressive increase in the rate of activated binding bonds as time advances. Specifically, at T/4, the percentage is 4.35%, which increases to 6.52% at T/2, 7.66% at 3T/4, and finally reaches 8.05% at the end of the simulation period. As concluded from this figure, the majority of the platelets near the surface, i.e., more than 90%, are inactive, resulting in random spots of platelets adhered to the surface, as shown in Fig. 11. The activation increases as time proceeds, but the rate of increases slows down at a later time.

Compared to the collagen-coated surface, the rate of platelet bond activation is lower. Furthermore, as depicted in Figure 11, the formed clots are smaller and more dispersed. This dispersion results in a more even distribution of platelets, preventing the formation of large clots. Over time, the percentage of activated bonds reaches a threshold, indicating a limit to the activation process under these conditions.This behavior can be crucial for understanding the dynamics of platelet activation and clot formation in different environments, highlighting the importance of surface properties in regulating these biological processes.

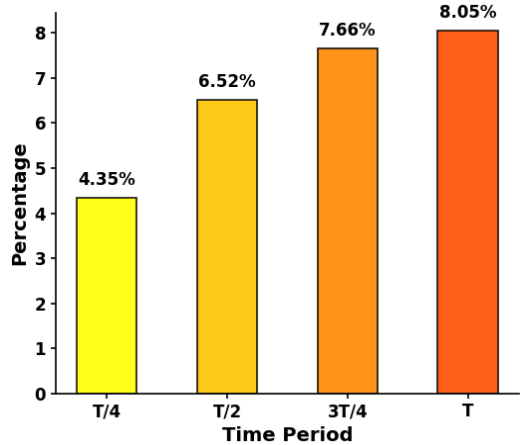

**Fig. 11:** Percentage of activated binding bonds with time.

Fig. 12 illustrates the velocity distribution function of platelets adhered to collagen at a shear rate of 500 1/s.. The shear stress of the flow is obtained as a post-processing of the DPD results. We obtained the velocity of the flow and then took a gradient of the velocity to compute shear stress. So, shear stress was not implemented in the code, and it is computed after the results are post-processed. As shown, most platelets exhibit velocities close to zero. This observation is consistent across four different simulations, all displaying a similar pattern. This similarity across simulations indicates a robust and predictable behavior of platelet adhesion in the presence of collagen. The low velocities suggest that once platelets adhere to collagen, they remain relatively stationary, which is crucial for forming stable thrombi. The consistency across simulations also implies that the model accurately captures the dynamics of platelet adhesion and can be reliably used for further studies.

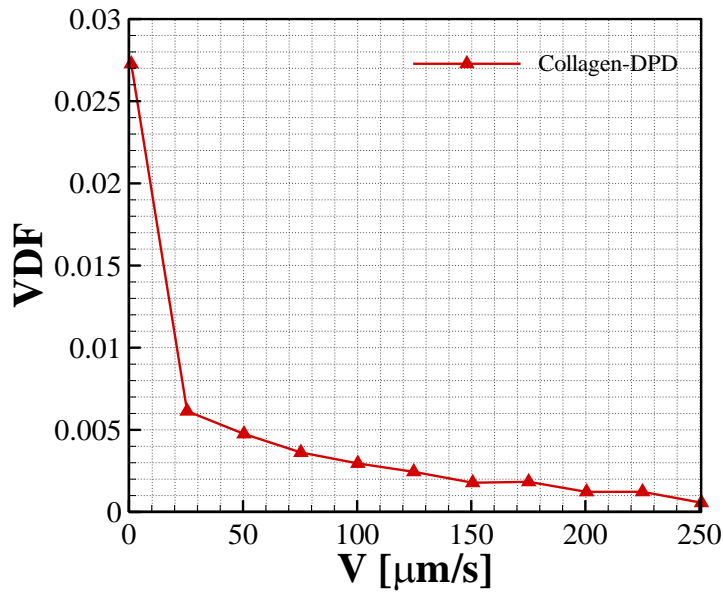

**Fig. 12:** Velocity distribution function of platelets over a surface covered by collagen at various conditions, shear rate: 500 1/s.

## 3.7 Platelet-vWF adhesion

The frames in Fig. 13 provide the DPD predictions for clot formation on a surface covered with vWF. The DPD pictures show that the formed clot is smaller than those on collagen. Note that the same numerical solver was utilized, but using different settings in Bell's law resulted in such a difference.

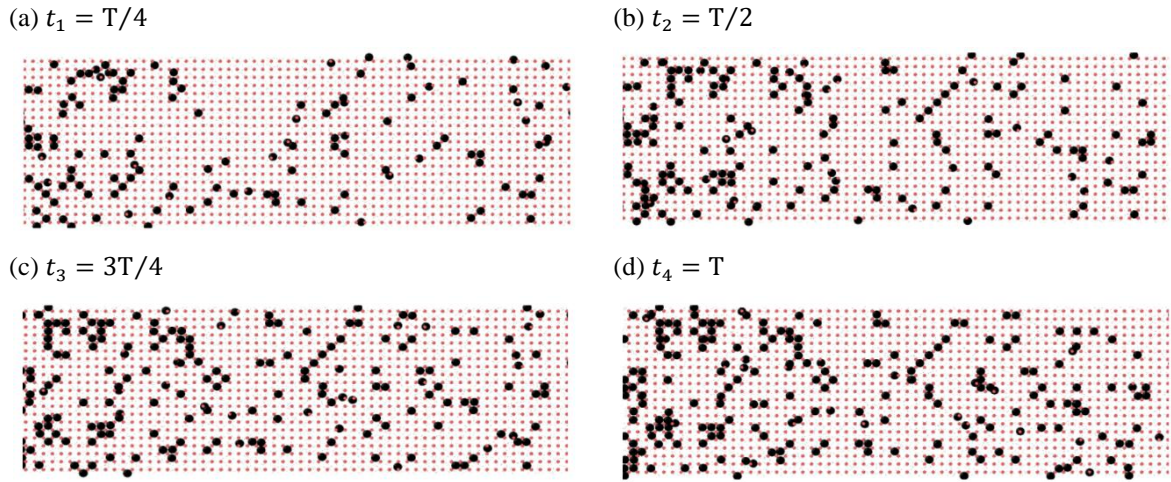

(a) $t_1 = T/4$  (b) $t_2 = T/2$
(c) $t_3 = 3T/4$  (d) $t_4 = T$

**Fig. 13:** Platelet adhesion to the VWF-coated surface simulated using the DPD method. The parameters used in the simulation are $k_0 = 1000, F_0 = 1, a = 1, a_t = 0, a_0 = 1, r_c = 0.67, T = 120$ s.

Fig. 14 illustrates the velocity distribution function (VDF) of platelets adhered to vWF. This figure shows that platelets adhered to vWF exhibit a broader range of velocities, including higher values significantly above zero. This suggests that, unlike collagen, platelets on vWF retain some degree of motion even after adhesion. The higher velocities observed indicate that platelets on vWF are not immobilized, allowing them to move slightly across the surface.

This difference in behavior is critical for understanding the distinct roles of vWF and collagen in hemostasis and thrombosis. vWF, while promoting platelet adhesion, does not bind platelets as tightly as collagen. This looser binding can be advantageous in certain physiological contexts where a balance between adhesion and movement is necessary. For instance, in the initial stages of clot formation, the ability of platelets to move and interact with other platelets and the vessel wall can enhance the overall clotting response. Moreover, the slight motion of platelets on vWF could facilitate the coverage of larger surface areas, promoting a more extensive interaction with the damaged vessel wall and other circulating cells. This dynamic adhesion process is essential for forming a hemostatic plug that is effective and adaptable to varying physiological conditions.

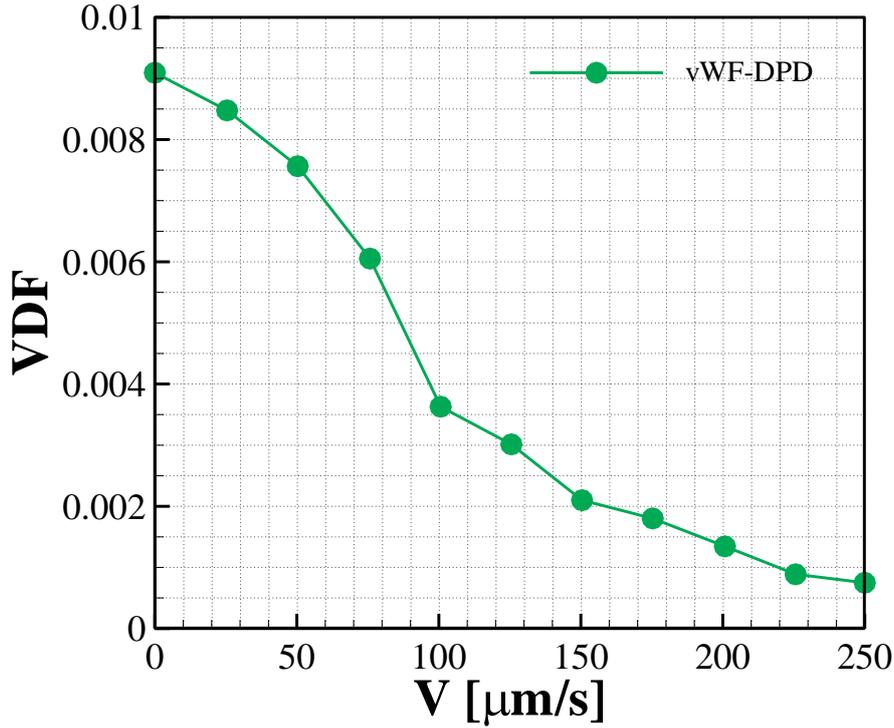

**Fig. 14:** Velocity distribution function of platelets over a surface covered by vWF at various conditions, shear rate: 500 1/s.

## 4. Discussion

Considering the importance of the platelet's role in hemostasis, the current work studied the thrombus formation process utilizing the DPD method equipped with probabilistic Bell's law and a time-dependent deterministic elastic force coefficient between the platelet and protein-coated surface. In the two-stage scheme designed to simulate platelet adhesion and aggregation, each activation stage plays a critical role in the overall process of clot formation. This model divides platelet activation into two distinct phases:

1. **Initial Activation**: This stage is characterized by activating platelets that are not yet bound to other cells or surfaces. Exposure to collagen and vWF triggers activation at this stage. The interaction with these substances initiates the platelets' adhesive capabilities, preparing them for subsequent interactions that lead to clot formation.
2. **Final Activation**: After the initial activation, platelets undergo a final transformation when they become fully active. This stage is marked by the activation of GPIIb/IIIa receptors, which are crucial for platelet aggregation. Once these receptors are activated, the platelets can bind more firmly to each other and to other components of the blood matrix. At this stage, the platelets are covered by a fibrin net, further stabilizing the aggregate and forming a robust clot structure.

This time-dependent approach was integral in delineating and replicating the intricate behaviors of platelets within the biological context, emphasizing the sequential progression from initial to full activation during clot formation. This methodological foundation facilitated the development and validation of our models using DPD within the LAMMPS simulation software framework. This approach uniquely integrated stochastic elements of Bell's law with a deterministic model of the elastic forces between platelets and a protein-coated surface, specifically addressing the interactions involving bonds with collagen and vWF.

Combining a deterministic elastic force model with a time-dependent coefficient and Bell's law permitted the current research to capture clotting behavior over a protein-coated surface with suitable accuracy when compared to our experiments. To the best of our knowledge, it is the first time that a numerical model can replicate in-vitro results for platelet-collagen adhesion. Note that the mechanism of platelet-collagen adhesion is complex[36]. The first numerical models of platelet-collagen adhesion were based on a deterministic elastic force model, i.e.,[13,14]. Later, Kaneva et al.[17] introduced a baseline deterministic model equipped with a complex statistical model. We employed a relatively straightforward statistical model in the current study that leverages Bell's law[20]. This approach was complemented by incorporating a time-dependent coefficient for the deterministic elastic force within the Dissipative Particle Dynamics (DPD) framework. This methodological choice proved to be particularly advantageous, enabling us to effectively simulate and reproduce the dynamic behaviors and patterns observed in biological systems in vivo.

Adopting statistical Bell's law, a principle well-regarded for its ability to model the random interactions between particles under varying forces, allowed us to lay a robust foundation for our simulations. The introduction of a time-dependent coefficient, a novel aspect of our approach compared to Ref. [14], further refined the model by accommodating changes in the elastic force of the system over time. This dynamic aspect of the model is crucial, as it mirrors the natural variability and adaptive responses observed in biological tissues, where mechanical properties can evolve in response to environmental or internal biological factors.

Our model's ability to mimic in vivo patterns not only substantiates the validity of using simplified statistical models in complex biological simulations but also highlights the potential of the DPD framework to be extended beyond traditional applications. By integrating time-dependent aspects into the modeling of elastic forces, we were able to capture a more accurate representation of the physiological processes, which are inherently time-variable.

These findings underscore the significance of our approach to advancing the field of computational biology. They suggest that even with simplified models when augmented with carefully considered, realistic dynamics, it is possible to achieve high fidelity in the simulation of biological systems. This has important implications for future research, particularly in developing predictive models of biological behavior that can be used in medical diagnostics, treatment planning, and the design of biomaterials.

Further studies should explore the scalability of this model to other biological systems and its integration with more complex computational frameworks. Investigating the impacts of varying the parameters of Bell's law and the time-dependent coefficients in different biological contexts could also yield deeper insights into the robustness and applicability of our approach. This could potentially lead to broader applications in computational tissue engineering and regenerative medicine, where understanding and predicting the mechanical behavior of biological systems over time is crucial.

**Concluding Remarks**

In this work, Bell's law governing the rate of binding of platelet receptors and ligands was implemented in the DPD framework. This allowed for a numerical approach to studying platelet adhesion behavior, akin to in vitro observations. This unique approach, where platelets adhere to surfaces fully coated with either collagen or vWF, has produced simulation results of random adhesion patterns. These patterns are quantitatively consistent with experimental data and demonstrate the feasibility of the DPD model in modeling platelet adhesion on both collagen and vWF. This advancement not only deepens our understanding of platelet dynamics but also enhances our ability to predict and study clotting behavior under various physiological conditions, opening the door for further research in thrombosis and hemostasis.

**Acknowledgments**
Ehsan Roohi would like to acknowledge Professor Zhongjun J. Wu from the University of Maryland School of Medicine for his valuable guidance and insightful support in this research. He also expresses his gratitude to Dr. Dong Han for his helpful discussions and assistance in this work. Additionally, the authors would like to sincerely thank Dr. Milad Ashrafizadeh for his valuable support during this study.

**Funding statement**
This research received no specific grant from any funding agency, commercial or not-for-profit sectors.

**Competing Interests**
The authors affirm that they have no conflict of interest regarding the subject matter or materials discussed in this manuscript. This statement underscores our commitment to maintaining the highest ethical standards in our research, ensuring that our interests have not influenced the results or interpretations presented in this study.

**Data availability statement**

The data that support the findings of this study are available from the corresponding author upon reasonable request.

**Author ORCID**

E. Roohi, https://orcid.org/0000-0001-5739-3210; A. Lotfian, https://orcid.org/0000-0002-1075-4717;

**CRediT authorship contribution statement**

Ali Lotfian: Schematic diagram, Software, Data Analysis, Investigation, Writing – review & editing. Ehsan Roohi: Conceptualization, Methodology, Supervision, Formal analysis, Data Analysis, Investigation, Writing – original draft, Writing – review & editing, Project administration, Funding acquisition.

**Declaration of competing interest**

The authors affirm that they have no conflict of interest regarding the subject matter or materials discussed in this manuscript. This statement underscores our commitment to maintaining the highest ethical standards in our research, ensuring that our interests have not influenced the results or interpretations presented in this study.

**Data availability statement**

Data will be available upon reasonable request to the corresponding author.

# Appendix I: Updated Platelet-Colagen Adhesion

We adjusted the time-dependent force model introduced by Eq. (10) in the body of the paper as follows:

$$a = \begin{cases} 0 & ; q_{ij} = 0 \\ a_t t + a_0 & ; q_{ij} > 0, t < t_{max} \\ a_{max} & ; q_{ij} > 0, t > t_{max} \end{cases} \qquad (A.I-1)$$

**Table A.I-1:** Force coefficients for Platelet-Clot.

| $a_t$ | $a_0$ | $r_a/r_c$ | $t_{max}$ | $\gamma$ | $r_c$ |
|---|---|---|---|---|---|
| 1 | 900 | 2/3 | 20 | 4.5 | 1 |

$t_{max}$ denotes the time required to reach maximum activation. Video 1. illustrates the adhesion dynamics of platelets to a collagen-covered surface over time under the conditions shown in Table A.I.1. The black circles in the video represent platelets, which randomly adhere to the surface. As time progresses, these randomly distributed adhesion sites expand, leading to the formation of larger adhesion regions. The clots formed by this model are more controlled than those obtained by the model in the body of the paper, as the force is not increasing unboundedly. Therefore, the threshold-based time-dependent model suggested in this appendix is more suitable than the model suggested in the body of the paper.

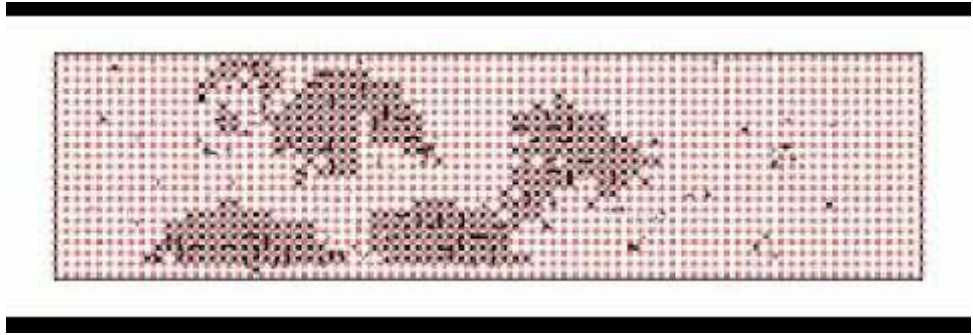

**Video 1:** Platelet adhesion to a collagen-covered surface under the conditions shown in Table A.I.1.
Link: https://www.youtube.com/watch?v=xElNUuOJHV8.